# Geiger Mode APD performance in a cryogenic two-phase Ar avalanche detector based on THGEMs

A. Bondar [a], A. Buzulutskov [a,*], A. Grebenuk [a], A. Sokolov [a],
D. Akimov [b], I. Alexandrov [b], A. Breskin [c]

[a] Budker Institute of Nuclear Physics, 630090 Novosibirsk, Russia
[b] Institute of Theoretical and Experimental Physics, 117218 Moscow, Russia
[c] Weizmann Institute of Science, 76100 Rehovot, Israel

**Abstract**

Characteristic properties of a Geiger Mode APD (G-APD) in a THGEM-based cryogenic two-phase Ar avalanche detector were studied in view of potential applications in rare-event experiments. G-APD signal amplitude and noise characteristics at cryogenic temperatures turned out to be superior to those at room temperature. The effective detection of avalanche scintillations from THGEM-multiplier holes in two-phase Ar has been demonstrated using a G-APD without wavelength shifter. At an avalanche gain of 60, the avalanche scintillation yield measured by the G-APD was as high as 0.9 photoelectrons per avalanche electron, extrapolated to $4\pi$ acceptance.

## 1. Introduction

In the concept of cryogenic two-phase avalanche detectors [1,2] the two-phase electron emission detector [3] is operated in an electron-avalanching mode using Gas Electron Multipliers (GEMs) [4] or thick GEMs (THGEMs) [5]. The development of such detectors is motivated by applications in rare-event experiments and in medical imaging: in coherent neutrino-nucleus scattering [6], dark matter search [7], solar and large scale neutrino detectors [8,9], and in Positron Emission Tomography [2]. In particular for the former application, a low-noise self-triggering detector is required with ultimate sensitivity, operated in single-electron counting mode. Most promising results were obtained with two-phase Ar avalanche detectors providing gains reaching $10^4$ with GEMs [10,11,12] and gains reaching $3\times10^3$ with THGEMs [13].

In this work we study a novel technique of signal recording in two-phase avalanche detectors, namely an optical readout of avalanche-induced scintillation photons from THGEM holes with Geiger Mode APDs (G-APDs, [14]). In detectors requiring ultimate sensitivities the optical readout, as compared to the charge readout, might be preferable in terms of overall gain and noise. In addition, the G-APD performance at low temperatures is expected to be superior to that at room temperature; however, existing data in such conditions vary in the literature [15,16,17,18,19,20], which motivated our new investigations.

Earlier studies of optical readout from THGEM in two-phase Ar [21] used G-APDs coated with a wavelength shifter (WLS), sensitive to VUV scintillation. In contrast, in the current work no WLS was used. This was based on the observation that Ar has effective avalanche-induced scintillations in the near infrared (NIR) [22], where the photon detection efficiency (PDE) of the G-APD can reach 20%. We present the first results on the optical readout from a THGEM multiplier in two-phase Ar with an uncoated G-APD; more elaborated results will be presented elsewhere [23].

## 2. Experimental setup

The experimental setups and procedures for investigating our first generation of two-phase avalanche detectors were already described in [10,11,12,13]. A new cryogenic chamber was recently assembled, with a larger volume (9 l), larger X-ray windows and better temperature stability (Fig. 1). In the two-phase mode, the chamber comprised a cathode mesh at the bottom, immersed in a 1 cm thick liquid Ar layer and a $25 \times 25$ mm$^2$ double-THGEM assembly placed in the saturated vapor above the liquid. The Ar was purified by an Oxisorb filter, providing an electron life-time of >20 μs in the two-phase mode. The THGEM geometrical parameters were the same as in [13].

* Corresponding author. Tel.: +7-383-3294833; fax: +7-383-3307163.
  *E-mail address:* a.f.buzulutskov@inp.nsk.su.

A G-APD (MRS APD "CPTA 149-35", [24]) was placed at a distance of 4 mm behind the second THGEM. It was optimized for the green-red range; it had a 4.41 mm$^2$ active area, 1764 pixels, a capacitance of 150 pF and a PDE of ~15% at 800 nm [24,25]. The quenching resistor of each pixel was measured to be 25 MΩ and 5.7 GΩ at room temperature and at 88 K respectively.

The signal from the G-APD was read out via a 1 m long twisted-pair cable connected to a fast amplifier (CPTA, [24]) with 300 MHz bandwidth and an amplification factor of 30. The ionization (anode) signal was read out from the last electrode of the second THGEM using a charge-sensitive amplifier with a shaping time of either 0.5 or 3 μs.

When studying the G-APD characteristics, the THGEM multiplier was inactive. In this case, at room temperature, the G-APD single-pixel amplitude was measured using the noise-amplitude spectrum. At cryogenic temperatures, however, the G-APD noise rate was too low. For this latter case the G-APD single-pixel amplitude, as well as the relative PDE, was measured using the amplitude spectrum of Ar scintillation light-pulses produced by a pulsed X-ray tube (with a frequency of 200 Hz) in the gas gap between the second THGEM and the G-APD.

For studying the two-phase avalanche detector performance with THGEM + G-APD, the signals were induced by 60 keV X-rays from an $^{241}$Am source.

## 3. Results

In Fig. 2 the G-APD pulse shapes of noise signals are compared at room and cryogenic temperatures, i.e. at 295 and 87 K: the single-, double- and triple-pixel signals are well identified. It is interesting to note that the pulse-shape is basically independent of the temperature, in contrast to observations in [16]. We did not observe any indications of after-pulses, in particular at cryogenic temperatures, for time scales of up to 10 ms.

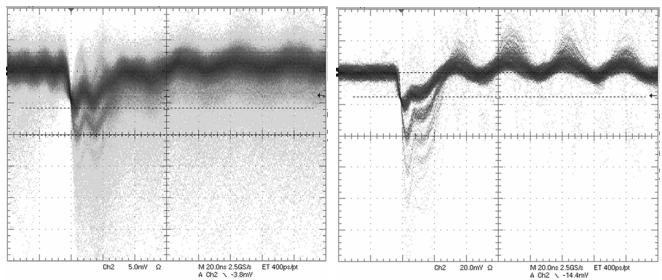

Fig. 2. G-APD noise signals in gain saturation mode: at 295 K (left, 5 mV/div, V=41 V) and 87 K (right, 20 mV/div, V=46 V). The time scale is 20 ns/div.

The major part of the signal had a width of 20 ns, reflecting the time structure of the Geiger discharge in the pixel. In addition the signal had a longer opposite-polarity tail, reflecting the characteristic response of the fast amplifier. The original bipolar pulses were transformed to unipolar, with a shaping time of 100 ns, the area of which provided the amplitude of the G-APD signals.

Fig. 3 shows the G-APD amplitude noise spectrum at 87 K and at a bias voltage of 44 V. The single-, double- and triple-pixel amplitude signals are well separated, providing the single-pixel amplitude to be well defined. From this spectrum one can estimate the cross-talk between pixels at this operation voltage: its value averaged over several measurement runs was about 40%. Accordingly, in the following the photoelectron yield is corrected for cross-talks by dividing by a factor of 1.4.

Fig. 4 shows the G-APD gain-voltage characteristics, namely the single pixel charge as a function of the bias voltage, at different temperatures. Here the over-voltage is defined as the difference between the bias and the breakdown voltage, the latter being defined at the intersection of the gain-voltage characteristic with the abscissa. One can see that the typical linear growth of the pixel amplitude ends by its saturation, at over-voltage of 7 V and 14 V at room and cryogenic temperatures respectively. Moreover, the maximum pixel amplitude (in the saturation mode) at cryogenic temperatures is 4-fold larger compared to room temperature (also seen in Fig. 2). The saturation effect observed here cannot be ascribed to the increase of the noise rate at higher voltages, since at cryogenic temperatures the noise rate is rather low (see Fig. 5).

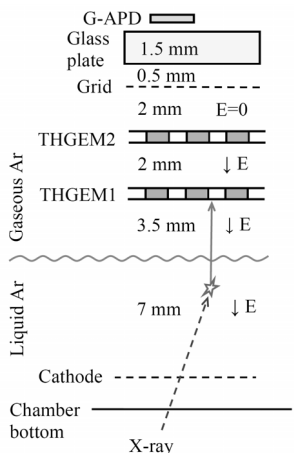

Fig. 1. Experimental setup.

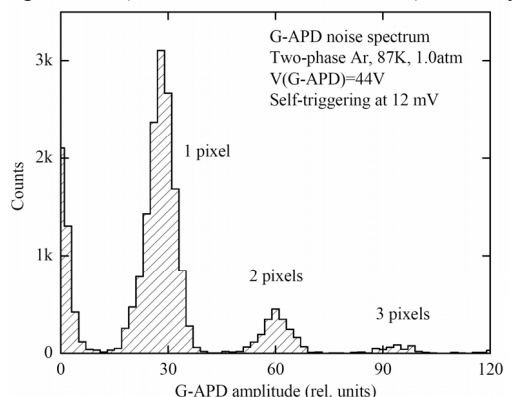

Fig. 3. G-APD amplitude noise spectrum at 87 K at the bias voltage of 44 V.

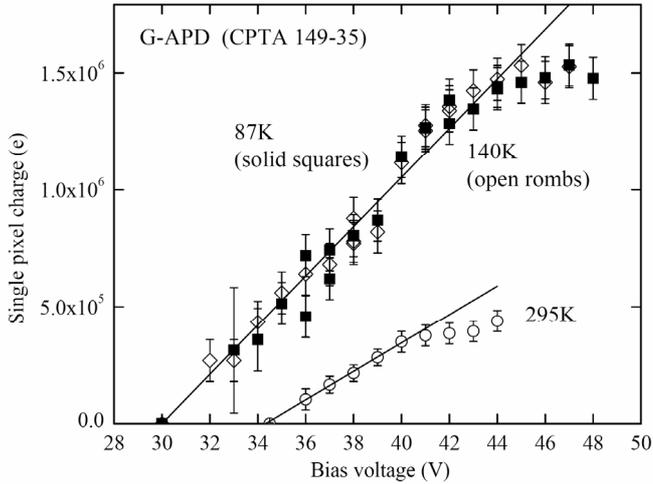

Fig. 4. G-APD single pixel charge as a function of the bias voltage at different temperatures: at 87 K, 140 K and 295 K.

As expected [17], compared to room temperature the breakdown voltage is decreased at cryogenic temperatures (Fig. 4): from 34.5 to 30 V respectively. On the other hand, the amplitude characteristics at 87 and 140 K turned out to be practically identical, though the noise rates are different (see Fig. 5).

Fig. 5 confirms the significant differences in G-APD performance between room temperature and cryogenic temperatures. The noise rate was measured by counting the noise pulses in a fixed time interval, the noise signals being recognized by their characteristic pulse-shape (Fig. 2). As expected, the noise rate considerably decreases with the temperature decreases: at 87 K it can be as low as 1 Hz at over-voltage of 8 V. Furthermore, the data points are described by different functions, indicating different mechanisms for the charge carrier generation: we used a linear function for the room temperature data, and an exponential function for the data at cryogenic temperatures.

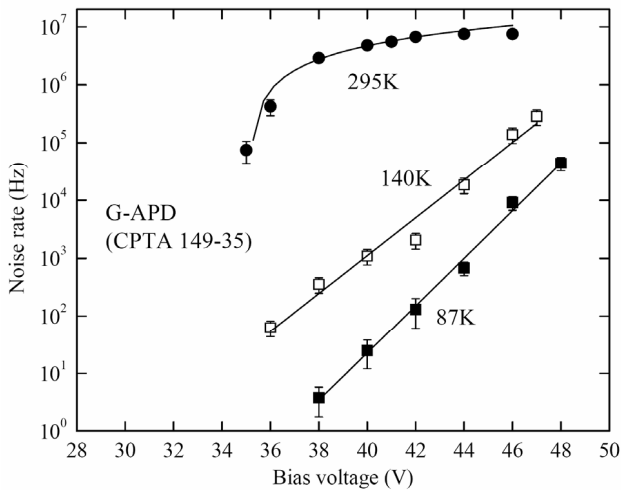

Fig. 5. G-APD noise rate as a function of the bias voltage at different temperatures. The fits to the data points are performed with an exponential function for the 87 and 140 K data, and with a linear function for the 295 K data.

Fig. 6 illustrates the dependence of the G-APD relative PDE on the bias voltage at 87 K; it shows the value proportional to the PDE, namely the G-APD average photoelectron number per scintillation pulse, induced by the X-ray pulse, recorded within a certain time gate. The PDE reaches a plateau at over-voltage of 5 V; a similar behavior was observed at 140 K.

We assume that the effects observed in the present work indicate more complicated mechanisms for the G-APD performance at cryogenic temperatures than what is expected from earlier studies [16,17,18]; but the particular mechanism may be strongly dependent on the G-APD structure.

We have observed avalanche scintillations in a THGEM-based two-phase Ar avalanche detector using a G-APD without WLS, i.e. insensitive to UV. Consequently, the scintillations most probably occurred in the NIR as discussed in section 1. As example, a scintillation and an ionization signal, induced by a 60 keV X-ray, are presented in Fig. 7, at a double-THGEM gain of 60 and a G-APD voltage of 44 V. The fast G-APD signal provides an effective means to study the electron emission and avalanche mechanisms in two-phase Ar. Its time structure reflects the electron emission processes at the liquid-gas interface [26]: the pulse spike at the beginning is induced by the fast electron emission component and the tail by the slow component (see details in [23]).

At this particular solid angle and avalanche gain (~60) the average number of photoelectrons (pe), recorded by the G-APD and corrected for cross-talks, was about 130 pe for a 60 keV X-ray converted in liquid Ar, producing about 900 initial electrons in the gas phase (prior to multiplication in the THGEM). That means that for the effective operation in a single electron counting mode (> 3 pe per initial electron), the solid angle, the THGEM gain and the light yield should be jointly increased by a factor of 20, which is possible to access.

The avalanche scintillation light yield can be better estimated from Fig. 8, showing the G-APD-to-THGEM amplitude ratio distribution; here the amplitudes are expressed in photoelectrons (pe) and avalanche electrons (e) respectively. The distribution average amounts to 0.0034 pe/e.

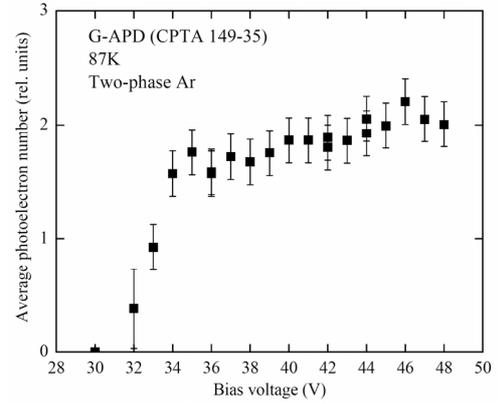

Fig. 6. The G-APD relative photon detection efficiency (PDE) as a function of the bias voltage at 87 K. Shown is the average photoelectron number per scintillation pulse, induced by the X-ray pulse, recorded within a 600 ns gate.

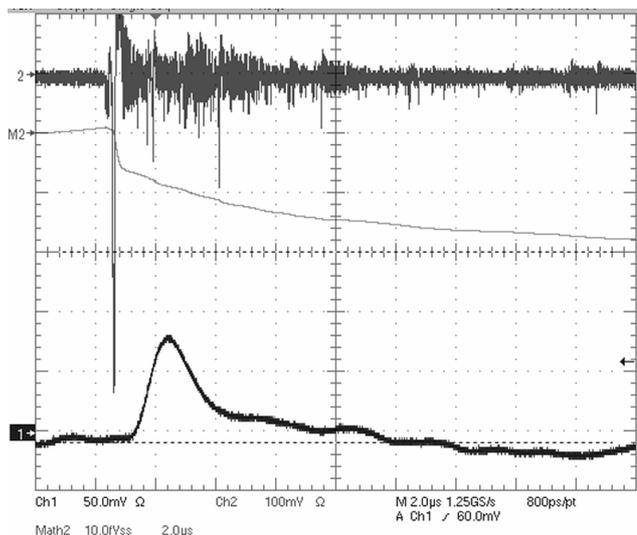

Fig. 7. Typical G-APD (upper trace) and THGEM (lower trace) signals induced by a 60 keV X-ray in two-phase Ar at 87 K and 1.0 atm, at a double-THGEM gain of 60, a G-APD bias voltage of 44 V, and an electrical field within the liquid Ar of 1.8 kV/cm. The vertical scale is 100 and 50 mV/div, respectively. The time scale is 2 µs/div. The THGEM amplifier shaping time is 0.5 µs. The integrated G-APD amplitude (middle trace) is 560 photoelectrons.

Taking into account the G-APD average solid angle with respect to the second THGEM, $\Delta\Omega/4\pi=2.7\times10^{-3}$, and correcting for cross-talks, we obtained the following G-APD yield extrapolated to $4\pi$ acceptance: 0.9 pe/e or, accounting for the G-APD PDE of 15% at 800 nm, 6 photons/e. This is a rather high yield; it should be compared to the value of ~1 photon/e presented in [22] for avalanche scintillations in Ar in the NIR region.

## 4. Conclusions

A novel optical concept of signal recording in two-phase avalanche detectors, with a G-APD measuring the THGEM avalanche scintillation photons, has been studied in view of potential applications in rare-event experiments.

The G-APD amplitude and noise characteristics at cryogenic temperatures turned out to be superior to those at room temperature; in particular at 87 K, the noise rates were of few Hz at the PDE efficiency plateau and the maximum G-APD gain was higher by a factor 4 compared to room temperature operation.

The effective detection of avalanche scintillations from THGEM multiplier holes in two-phase Ar has been demonstrated with G-APDs not coated with a WLS. At an avalanche gain of 60, the THGEM + G-APD yielded about 130 photoelectrons per 60 keV X-ray converted in liquid Ar. The avalanche scintillation yield measured by the G-APD was 0.9 photoelectrons per avalanche electron, extrapolated to $4\pi$ acceptance.

In practice, the optical readout of two-phase detectors would comprise a matrix of G-APDs placed behind the THGEM multiplier with a pitch of ~1 cm, to cover the active area with a spatial resolution sufficient for rare-event experiments. For example, for a 100 kg liquid Ar TPC with a volume of $40\times40\times40$ cm$^3$ the total number of G-APDs would amount to 1600, which is not too much. Such a THGEM + G-APD assembly would be robust, relatively cheap and would have good performance. Further studies are in progress.

We are grateful to Y. Tikhonov for the support, R. Snopkov and A. Chegodaev for the development of the experimental setup, M. Danilov and E. Kravchenko for providing the G-APDs, A. Akindinov, J. Haba and Y. Kudenko for the discussion of the results. This work was supported in part by RFFI grant 09-02-12217-ofi_m and by Federal Special Program "Scientific and Scientific-Educational Specialists of Innovation Russia" in 2009-2013.

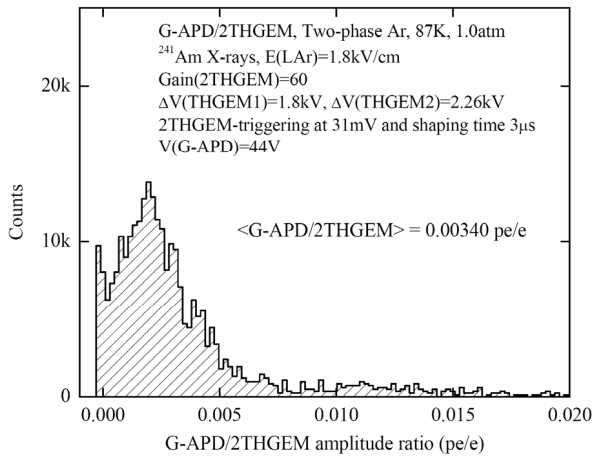

Fig. 8. G-APD to THGEM amplitude ratio distribution induced by 60 keV X-rays in two-phase Ar at 87 K and 1.0 atm, at double-THGEM gain of 60, G-APD bias voltage of 44 V and electrical field within liquid Ar of 1.8 kV/cm. The G-APD and THGEM amplitudes are expressed in photoelectrons and (avalanche) electrons, respectively.